\documentclass
[10pt,twocolumn,showpacs,preprintnumbers,amsmath,amssymb]{revtex4}%
\usepackage{graphicx}
\usepackage{bm}
\usepackage{graphicx,amsmath,mathrsfs}
\usepackage{amsmath}
\usepackage{amsfonts}
\usepackage{amssymb}%
\providecommand{\U}[1]{\protect\rule{.1in}{.1in}}
\input epsf.sty

\begin{document}
\title{A finite range pairing force for density functional theory in superfluid nuclei}
\author{Y. Tian$^{1,2}$, Z.Y. Ma$^{1}$, P. Ring$^{2}$}
\affiliation{(1)~China Institute of Atom Energy, Beijing 102413, China}
\affiliation{(2)~Physikdepartment, Technische Universit\"at M\"unchen, D-85748, Garching, Germany}
\date{\today}

\begin{abstract}
The problem of pairing in the $^{1}$S$_{0}$ channel of finite nuclei
is revisited. In nuclear matter forces of separable form can be
adjusted to the bare nuclear force, to any phenomenological pairing
interaction such as the Gogny force or to exact solutions of the gap
equation. In finite nuclei, because of translational invariance, such
forces are no longer separable. Using well known techniques of Talmi
and Moshinsky we expand the matrix elements in a series of separable
terms, which converges quickly preserving translational invariance
and finite range. In this way the complicated problem of a cut-off at
large momenta or energies inherent in other separable or zero range
pairing forces is avoided. Applications in the framework of the
relativistic Hartree Bogoliubov approach show that the pairing
properties are depicted on almost the same footing as by the original
pairing interaction not only in nuclear matter, but also in finite
nuclei. This simple separable force can be easily applied for the
investigation of pairing properties in nuclei far from stability as
well as for further investigations going beyond mean field theory
\end{abstract}

\maketitle



Along with improved techniques to investigate more precisely nuclear systems
being considered as well known, the recent generation of radioactive beam
facilities enables us to examine exotic systems with extreme isospin values.
Therefore experimental and theoretical studies of nuclei far from the valley
of $\beta$-stability are presently at the forefront of nuclear science.
Experiments with radioactive nuclear beams have already in the past discovered
a number of new structure phenomena in exotic nuclei with extreme isospin
values, and the next radioactive-beam facilities in construction will present
new exciting opportunities for the study of the nuclear many-body systems.

Nuclei far from stability have also an important influence on astrophysical
processes. Therefore the study of such nuclei has wide ranging applications in
modern nuclear astrophysics. Unfortunately many of the nuclei of interest in
this context have such large neutron excess, that it will be impossible in
near future and probably even excluded in far future to investigate them on
earth by experiments in the laboratory. Therefore it is extremely important to
provide a powerful theory for a reliable description of nuclei close to the
limits of stability. It should be based on a consistent treatment of both
ground and excited states and should allow for predictions of nuclear
properties in areas, which are hard or impossible to access by future
experiments. Ab initio calculations and multi-configuration mixing within the
shell-model are definitely a goal, but, so far, they can only be applied in
light nuclei. At present, for a universal description of nuclei all over the
periodic table, Density Functional Theory (DFT) based on the mean-field
concept provides a very reasonable concept. DFT has been introduced in the
sixties in atomic and molecular physics~\cite{HK.64,KS.65} and shortly after
that in nuclear physics under the name 'density dependent Hartree-Fock
theory'~\cite{VB.72,DG.80}. Today it is widely used for all kinds of quantum
mechanical many-body systems. DFT can, in principle, provide an exact
description of many-body dynamics, if the exact density functional is known,
but for systems such as nuclei one is far from a microscopic derivation and
the most successful applications determine the functional in a
phenomenological way. Starting from basic symmetries the parameters are
adjusted to characteristic experimental data in finite nuclei and nuclear
matter. In addition nuclei are self bound systems. One usually considers
densities in intrinsic frames and it is still under debate whether density
functional theory can be exact under these circumstances~\cite{Eng.07,Gir.08}.
Nonetheless in practice DFT provides in many nuclei all over the periodic
table an amazingly successful description of the complicated many-body
system~\cite{BHR.03,VALR.05}.

Conventional DFT with a functional $E[\rho]$ depending only on the single
particle density can be applied in nuclear physics practically only in a few
doubly closed shell nuclei. In all nuclei with open shells, and this is the
vast majority of all nuclei in the periodic table, the inclusion of
particle-particle ($pp$) correlations is essential for a correct description
of structure phenomena. Although, in principle the effective $pp$-interaction
is isospin dependent with a $T=0$ and a $T=1$ part, for the vast majority of
pairing effects in nuclei only the $T=1$ part is important. In fact, little is
known on the effective $T=0$ part. It is still an open question, whether the
effective $pp$-interaction in the $T=0$ channel is strong enough to produce a
pairing condensate with~\cite{WFS.71,Goo.79,SWy.97,MFC.00a,Afa.07}. We
therefore restrict ourselves in the following discussion to $T=1$ pairing
correlations between like particles.

In the framework of \ DFT pairing correlations are ta\-ken into account in the
form of Hartree-Bogoliubov theory, where the energy functional $E[\rho
,\kappa]$ depends not only on the normal density $\rho=\left\langle a^{+}%
a^{{}}\right\rangle $ but also on the pairing density $\kappa=\left\langle
a^{+}a^{+}\right\rangle $. Both densities can be combined to the so-called
Valatin density $\mathcal{R}$~\cite{Val.61} with the property $\mathcal{R}%
^{2}=\mathcal{R}$. This shows clearly that the Hartree-Bogoliubov version of
DFT is a generalized mean field theory. In Ref.~\cite{Kuc.89,KuR.91,Ring.96}
relativistic Hartree-Bogoliubov (RHB) theory has been introduced for the
treatment of pairing correlations in relativistic DFT. It turns out that
pairing itself is a non-relativistic effect influencing only the vicinity of
the Fermi surface. The effect of the pairing field on the small components can
be neglected to a very good approximation~\cite{SR.02}.

For nuclei close to the $\beta$-stability line, pairing has been included in
non-relativistic DFT~\cite{Vau.73} or in the relativistic DFT~\cite{GRT.90} in
the form of the simple constant gap approximation, where the pairing gap
$\Delta$ is obtained from odd-even mass differences. The occupation numbers
$v_{k}^{2}$ are then determined by the BCS-ansatz. The pairing tensor is
diagonal with the elements $\kappa_{k}=u_{k}v_{k}$ where $u_{k}^{2}+v_{k}%
^{2}=1$ and the pairing part of the density functional is given by the trace
of $\kappa$: $E_{pair}[\kappa]=-\Delta\sum_{k}\kappa_{k}$. Of course, this sum
diverges and therefore one has to restrict the sum over $k$ to a pairing
window. This introduces an additional parameter, which is not well determined
by experiment. Several prescriptions can be found in the literature to treat
the pairing window~\cite{MN.92,BFH.85}.

This way to include pairing correlations corresponds to the seniority
model~\cite{Ker.61}, which is also sometimes called monopole pairing, where
only pairs coupled to angular momentum $J=0$ feel the effective $pp$%
-interaction. Of course, this is a very simplified description of pairing
correlations in nuclei and therefore this force is often replaced by a zero
range force%
\begin{equation}
V^{pp}=V_{0}\delta(\mathbf{r}_{1}\mathbf{-r}_{2}){\scriptstyle\frac{1}{2}%
}(1-P^{\sigma})\label{E1}%
\end{equation}
which is sometimes chosen to be density dependent~\cite{BE.91}. For zero range
forces as in Eq. (1) the factor $\frac{1}{2}(1-P^{\sigma})$ projects on the
$^{1}S_{0}$ channel. This force is simple to handle in $r$-space and it is
therefore often used in non-relativistic HF-BCS-calculations with Skyrme
forces~\cite{BDF.90} and in relativistic density functionals based on
point-coupling models~\cite{BMM.02,NVR.06a,NVR.06b}. Unfortunately the
corresponding pairing energy diverges in this case too. As in the case of
seniority pairing one needs a pairing window. Sharp pairing windows have the
tendency to lead to flip-flop solutions in self-consistent theories and
therefore one uses in most of the present applications soft pairing
windows~\cite{BRR.00b} with rather arbitrary cut off parameters.


For nuclei far from stability the BCS approximation presents only a poor
approximation. In particular, in drip-line nuclei the Fermi level is found
close to the particle continuum. The lowest particle-hole ($ph$) or
particle-particle ($pp$) modes are often embedded in the continuum, and the
coupling between bound and continuum states has to be taken into account
explicitly. In these cases the BCS model does not provide a correct
description of the scattering of nucleonic pairs from bound states to the
positive energy continuum because several levels in the continuum become
partially occupied leading to a gas of nucleons surrounding the nucleus.
Including the system in a box of finite size leads to unreliable predictions
for nuclear radii depending on the size of this box. In the non-relativistic
case, it has been shown that the Hartree-Fock-Bogoliubov (HFB) theory in the
continuum provides a very elegant solution to this
problem~\cite{Bul.80,DFT.84,DNW.96} and this method has been applied also to
investigations of drip line and halo nuclei within covariant
DFT~\cite{MR.96,PVL.97}. The HFB theory presents a unified description of
$ph$- and $pp$-correlations~\cite{RS.80} on a mean-field level by using two
average potentials: the self-consistent Hartree-Fock field $\hat{\Gamma}$
which encloses all the long range $ph$-correlations, and a pairing field
$\hat{\Delta}$ which sums up the $pp$-correlations.

In order to avoid the complicated problems of a pairing cut-off Gogny derived
his energy functional~\cite{DG.80} from a finite range force of Brink-Booker
type. The finite range guarantees that the force decreases as a function of
the momentum transfer and the gap equation converges without any problems.
Thus a pairing cut off is not necessary. The parameters of this force have
been adjusted very carefully in a semi-phenomenological way by
Gogny~\cite{DG.80} and his collaborators~\cite{BGG.91,Cha.07} to
characteristic properties of the microscopic effective interactions and to
experimental data. Over the years this method has turned out to be a very
successful way to describe pairing correlations in nuclei and it is often used
as benchmark for more microscopic investigations~\cite{KRS.89,SRRi.02}.

Of course, mean field calculations with finite range forces turn out to
require a substantial numerical effort, in particular in three-dimensional
applications to triaxial nuclei~\cite{GG.83} and to rotating
systems~\cite{ERo.93,AKRE.00,ARK.00} or in applications going beyond mean
field such as the Generator Coordinate Method (GCM)~\cite{RER.04} connected
with projection to good angular momentum~\cite{NVR.06a,YMA.08b} and particle
number~\cite{RER.00,NVR.06b}. Since one needs nowadays systematic
investigations over a wide range of nuclei~\cite{DGG.06} it is highly
desirable to find an effective interaction in the pairing channel which is
numerically simpler without loosing the nice properties of this force. In this
investigations we propose a new realistic pairing force, which is carefully
adjusted in a separable form in momentum space to nuclear matter properties of
the Gogny force. It turns out that this force is simple enough, that its
matrix elements in finite nuclei can be expressed as a rather limited sum of
separable terms.


We start our investigations in symmetric nuclear matter. The effective pairing
force in the $pp$-channel can be represented by a sum of all diagrams
irreducible in $pp$-direction. In lowest order it corresponds to the bare $NN$
interaction. Recently Duguet~\cite{Dug.04} proposed a microscopic effective
interaction to treat pairing correlations in nuclear matter in the $^{1}S_{0}$
channel. Starting from an ansatz separable in momentum space, he derived after
several approximations an effective pairing force with zero range for
practical applications in the context of Skyrme calculations in $r$-space for
finite nuclei. We start from a similar ansatz in nuclear matter, however,
instead of reducing it to zero range, we transform the force obtained in this
way from momentum space to $r$-space. Then calculations in finite nuclei are
carried out in terms of an expansion in the eigenfunctions of a harmonic
oscillator. This is particularly simple for a Gaussian ansatz leading to
analytical expressions for the matrix elements. In Refs. \cite{DL.08,LDBM.09}
a similar technique was used based on spherical Bessel functions.

The wave functions in infinite nuclear matter are characterized by the
momentum ${{\mathbf{k}}}$ and the spin $s$
\begin{equation}
\varphi_{{{\mathbf{k}}}s}({{\mathbf{r}}})=e^{i{{\mathbf{k}}} \cdot
{{\mathbf{r}}}}\chi^{1/2}_{s}.
\end{equation}
The antisymmetric matrix element of the pairing interaction $V_{\mathrm{sep}%
}^{^{1}S_{0}}$ in the plane-wave basis has the form%
\begin{align}
&  \langle{{\mathbf{k}}}_{1}^{{}}s_{1}^{{}}{{\mathbf{k}}}_{2}^{{}}s_{2}^{{}%
}|{\scriptstyle\frac{1}{2}}(1-P_{\sigma})V^{^{1}S_{0}}|{{\mathbf{k}}}%
_{1}^{\prime}s_{1}^{\prime}{{\mathbf{k}}}_{2}^{\prime}s_{2}^{\prime}%
\rangle_{a}~~~~~~~~~~~~~~~~~~~~~~~~~~~\\
&  =\frac{1}{2}\langle{{\mathbf{k}}}_{1}^{{}}{{\mathbf{k}}}_{2}^{{}}%
|V^{^{1}S_{0}}|{{\mathbf{k}}}_{1}^{\prime}{{\mathbf{k}}}_{2}^{\prime}%
\rangle_{s}(\delta_{s_{1}^{{}}s_{1}^{\prime}}\delta_{s_{2}^{{}}s_{2}^{\prime}%
}-\delta_{s_{1}^{{}}s_{2}^{\prime}}\delta_{s_{2}^{{}}s_{2}^{\prime}}),
\end{align}
where
\begin{equation}
\langle{{\mathbf{k}}}_{1}^{{}}{{\mathbf{k}}}_{2}^{{}}|V^{^{1}S_{0}%
}|{{\mathbf{k}}}_{1}^{\prime}{{\mathbf{k}}}_{2}^{\prime}\rangle=\langle
{{\mathbf{k}}}|V^{^{1}S_{0}}|\mathbf{k}^{\prime}\rangle(2\pi)^{3}%
\delta({{\mathbf{K}}}-{{\mathbf{K}}}^{\prime}).
\end{equation}
${{\mathbf{K}}}\mathbf{=k}_{1}+\mathbf{k}_{2}$ and ${{\mathbf{k=}}}\frac{1}%
{2}{{\mathbf{(k}}}_{1}-{{\mathbf{k}}}_{2}{{\mathbf{)}}}$ are the total and
relative momentum of a particle pair, respectively. It has been pointed that
the bare interaction in the $^{1}S_{0}$ channel is to a good approximation
separable and nonlocal at low energy~\cite{BJ.76}. The center of mass part of
the matrix element is therefore approximated by a separable form,
\begin{equation}
\langle{{\mathbf{k}}}|V_{\mathrm{sep}}^{^{1}S_{0}}|{{\mathbf{k}}}^{\prime
}\rangle=-Gp(k)p(k^{\prime}). \label{sp}%
\end{equation}
The isospin quantum number is not specified in this expression, since the form
of the matrix elements in the ($T=1$) channel is trivial. $G$ is the strength
parameter of this pairing interaction.

In the $^{1}S_{0}$ channel we find the following gap equation in the plane
wave basis, the usual BCS equation~\cite{BCS.57a,BCS.57b},
\begin{equation}
\Delta(k)=-\int_{0}^{\infty}\frac{k^{\prime2}dk^{\prime}}{2\pi^{2}%
}v(k,k^{\prime})\frac{\Delta(k^{\prime})}{2E(k^{\prime})},\label{gapeq}%
\end{equation}
with the quasi-particle energy%
\begin{equation}
E(k)=\sqrt{(\epsilon(k)-\mu)^{2}+\Delta^{2}(k)}%
\end{equation}
and the in medium on-shell single particle energies $\epsilon(k)$ associated
with the state $\varphi_{{{\mathbf{k}}}}$. $\mu$ is the chemical potential
determined by the density and $v(k,k^{\prime})=-Gp(k)p(k^{\prime})$. For a
given pairing interaction, one can solve the BCS gap equation and calculate
the corresponding gaps as a function of the density, i.e. as a function of the
Fermi momentum in nuclear matter. This relationship between the gap at the
Fermi surface and the Fermi momentum determines the properties of the pairing correlations.

\begin{figure}[ptb]
\includegraphics[scale=0.25]{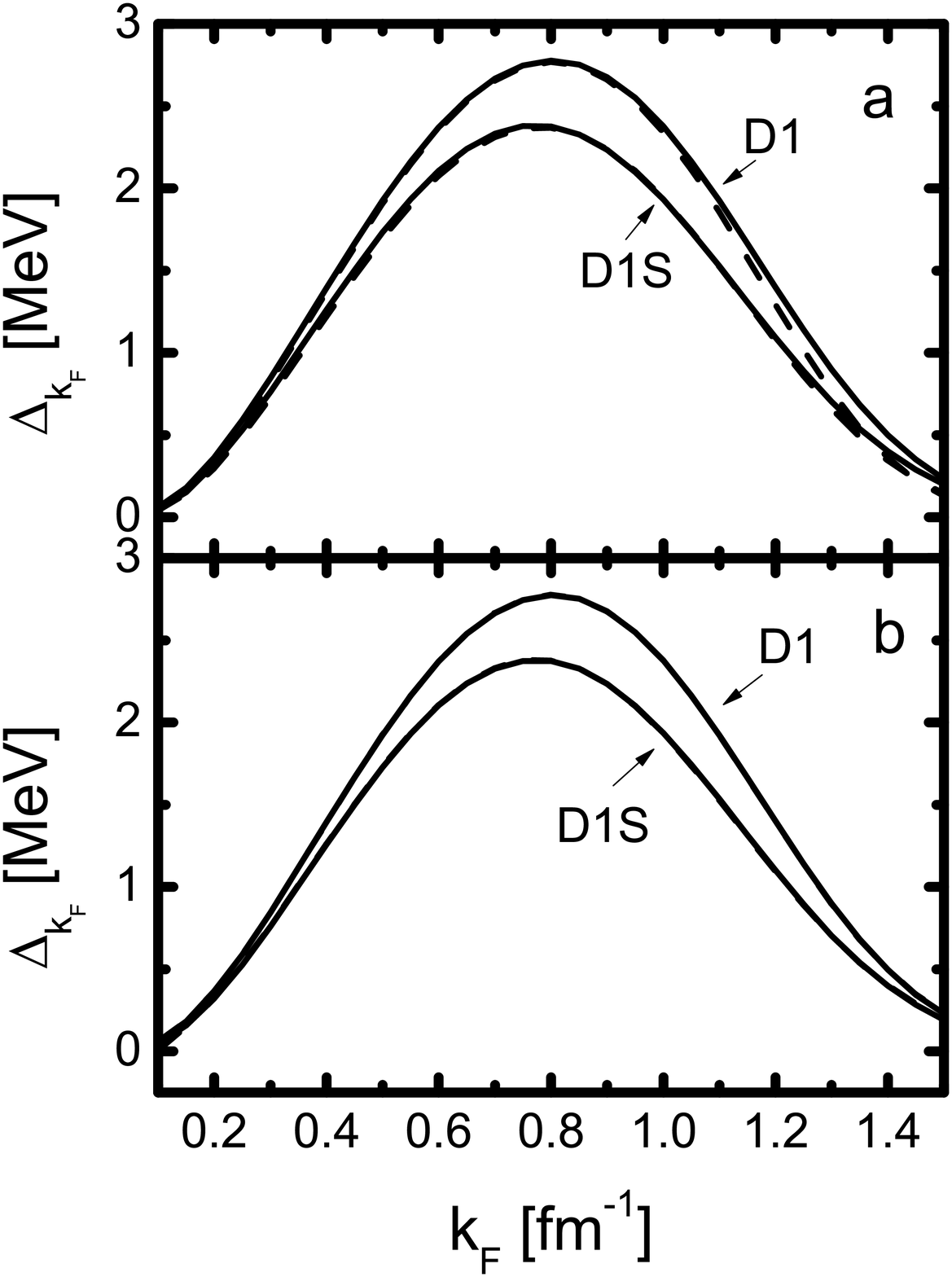}\caption{ Comparison of $^{1}S_{0}$
pairing gaps at the Fermi surface $\Delta(k_{F})$ as a function of the density
for the Gogny forces D1 and D1S (solid curves) and the corresponding separable
forces (dashed curves), where the function $p(k)$ in Eq.~(\ref{sp}) is
represented in panel (a) by one Gaussian in Eq. (\ref{gauss-r}) and in panel 3
by a sum of three Gaussians in Eq. (\ref{gauss3}).}%
\label{fig1}%
\end{figure}

Inserting the separable interaction $Gp(k)p(k^{\prime})$ into Eq.~(\ref{gapeq}%
), the solution of the gap equation is trivial $\Delta(k)=\Delta_{0}p(k)$,
where $\Delta_{0}$ is the gap at zero momentum satisfying the equation
\begin{equation}
1=\int_{0}^{\infty}\frac{k^{2}dk}{4\pi^{2}}\frac{Gp^{2}(k)}{\sqrt
{(\epsilon(k)-\mu)^{2}+\Delta_{0}^{2}p^{2}(k)}} \label{gapeq1}%
\end{equation}
and therefore depending on the Fermi momentum $k_{F}$. The gap at the Fermi
surface is obtained through%
\begin{equation}
\Delta(k_{F})=\Delta_{0}(k_{F})p(k_{F})
\end{equation}
which shows that the bell shaped curve $\Delta(k_{F})$ determines the form of
the functional dependence of the function $p(k)$ in Eq. (\ref{sp}).

\begin{figure}[ptb]
\includegraphics[scale=0.25]{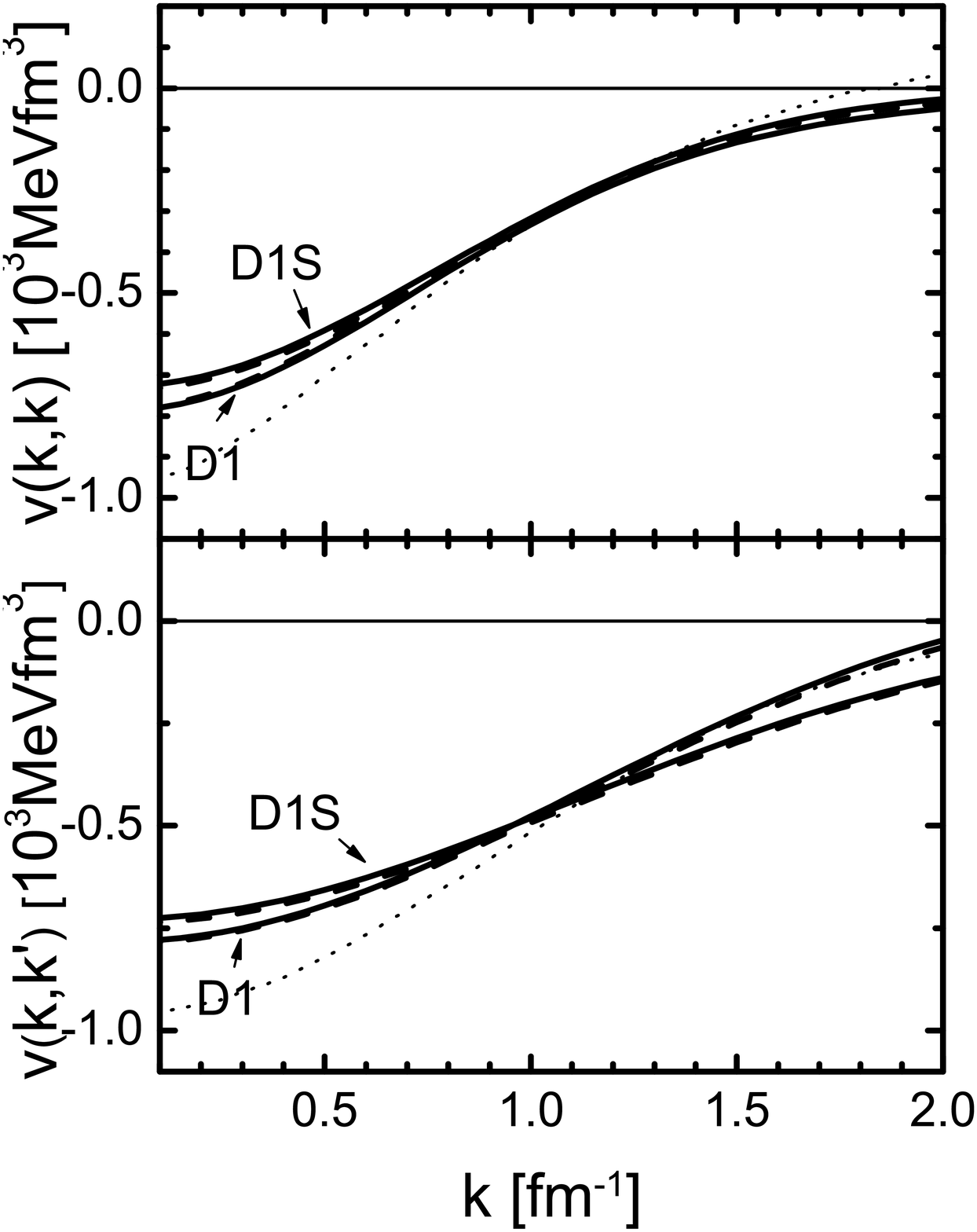}
\caption{Diagonal matrix elements (upper part) $v(k,k)$ and
non-diagonal matrix elements (lower part) $v(k,k^{\prime})$ with
$k^{\prime}=0.009$ fm$^{-1}$ as a function of the momentum $k$ in the
$^{1}S_{0}$ channel for Gogny force and the corresponding separable
forces. The dotted curve
represents $V_{{\protect\tiny \hbox{low~k}}}$.}%
\label{fig2}%
\end{figure}

For RHB calculations in finite nuclei it is common to use the Gogny
force~\cite{GEL.96} or a zero-range $\delta$-interaction
\cite{BMM.02}. We apply this method to derive a separable forces,
i.e. the function $p(k)$ in
Eq.~(\ref{sp}) by mimicking the non-relativistic Gogny force in the $^{1}%
S_{0}$ channel of nuclear matter. Of course this procedure depends on the self
energies $\epsilon(k)$. When we deal with the Gogny force this self energies
Eqs.~(\ref{gapeq}) and (\ref{gapeq1}) are no longer free single particle
energies. Medium corrections have to be included, using for instance the
Brueckner-Hartree-Fock approximation. For simplicity, however, we approximate
the self energy in this investigation in a phenomenological way by the single
particle energies of the form%
\begin{equation}
\epsilon(k)=V(\mu)+\sqrt{k^{2}+M^{\ast2}(\mu)},
\end{equation}
where the effective mass $M^{\ast}=M+S=M+g_{\sigma}\sigma$ and the effective
vector field $V=g_{\omega}\omega$ are given by the effective meson fields.
They are determined by the solution of the nonlinear Klein Gordon equations of
the RMF model with the parameter set NL3~\cite{NL3}.

First we solve the gap equation Eq.~(\ref{gapeq1}) in the $^{1}S_{0}$ channel
for nuclear matter at various densities with the Gogny force ~\cite{KRS.89}.
In order to determine a separable force, we first choose for the function
$p(k)$ the Gaussian ansatz%
\begin{equation}
p(k)=e^{-a^{2}k^{2}}. \label{gauss-k}%
\end{equation}
The two parameters $G$ and $a$ are fitted to the function $\Delta(k_{F})$.
Comparing with the Gogny force, we obtain two sets of parameters $G=738$
MeV$\cdot$fm$^{3}$ and $a=0.636$ fm for the parameter set D1~\cite{DG.80} and
$G=728$ MeV$\cdot$fm$^{3}$ and $a=0.644$ fm for the set D1S~\cite{D1S}. In the
upper part of Fig.~\ref{fig1} we show the pairing gaps in the $^{1}S_{0}$
channel obtained from the full Gogny forces and the corresponding separable
approximations (dashed curves). The separable forces reproduce the curves for
the gap almost perfectly, especially in the case of the parameterization D1S.
As we see the pairing gap can be very well reproduced at low densities by a
Gaussian shape and we observe only slight deviations for higher densities with
$k_{F}>1.0$ fm$^{-1}$. Of course, we could have chosen also some other form of
$p(k)$ to map the curves of the gap completely. In the lower part of
Fig.~\ref{fig1} we obtain perfect fits to the results of the Gogny force by
using a linear combination of three Gaussians with different widths and
amplitudes for $p(k)$.%
\begin{equation}
p(k)={\displaystyle\frac{1}{3}\sum\limits_{i=1}^{3}} e^{-a_{i}^{2}k^{2}}.
\label{gauss3}%
\end{equation}
with the parameters $G=731.25$ MeV$\cdot$fm$^{3}$ and $a_{1}=0.69$ fm,
$a_{2}=0.16$ fm, $a_{3}=0.47$ fm for the parameter set D1~\cite{DG.80} and
$G=742.09$ MeV$\cdot$fm$^{3}$ and $a_{1}=0.74$ fm, $a_{2}=0.26$ fm,
$a_{3}=0.36$ fm for the parameter set D1S \cite{D1S}. In the following
calculations, however, we apply only one Gaussian form for $p(k)$ that is much
simpler and easier to handle and that is good enough to describe the pairing
properties in the finite nuclei.

In order to investigate the behavior of the separable forces in more detail we
also plot in Fig.~\ref{fig2} the diagonal and non-diagonal matrix elements for
each of these cases. The matrix elements in the $^{1}S_{0}$ channel of the
separable forces obtained from Gogny D1 and D1S are very close to those of the
original Gogny forces. Similar results have been found in Refs.
\cite{SKM.03,Dug.04}. The Gogny force is an effective force in the nuclear
medium. In the case of bare nucleon-nucleon forces with a strong short range
repulsion this would be no longer the case. It is known~\cite{BKS.03} that, by
integrating out the high momentum components of the bare force, one obtains
from most of the realistic $NN$-potentials the same interaction a low momenta
$V_{{\tiny \hbox{low~k}}}$. This low momentum $NN$ interaction
$V_{{\tiny \hbox{low~k}}}$ can describe the two-nucleon system at low energy
very well. Although the matrix elements of various realistic bare $NN$
interactions are scattered, their low momentum part $V_{{\tiny \hbox{low~k}}}$
has the same shape. It has been found that the matrix element of the separable
form of AV18 is very close to that of $V_{{\tiny \hbox{low~k}}}$ obtained from
AV18 and very different from that of the potential AV18 itself. Therefore we
plot in Fig.~\ref{fig2} also the matrix elements of $V_{{\tiny \hbox{low~k}}}%
$. This illustrates the physical content of the separable force as an
effective interaction for the area of low energies. It is also shown in
Fig.~\ref{fig2} that the matrix elements of the Gogny force D1S have a very
similar behavior to that of $V_{{\tiny \hbox{low~k}}}$. This indicates that
the Gogny force has a clear link to the bare $NN$ force, especially for the
parameterization D1.


Now we turn to the solution of the RHB equations in finite nuclei. As
discussed earlier the separable force has simple Gaussian form and it is
fitted to reproduce the density dependence of the gap at the Fermi surface in
nuclear matter derived from the Gogny force. This force is definitely not
identical to the full Gogny force in the $^{1}S_{0}$-channel. Therefore, in
order to apply these separable pairing force in calculations of finite nuclei,
one has to study to what degree such a separable force can describe the
pairing properties of these systems. For that purpose we use RHB theory in
several spherical isotope chains, e.g. in Sn- and in Pb-isotopes.

First, we transform the force (\ref{sp}) from the momentum space to the
coordinate space and obtain
\begin{equation}
V(\mathbf{r}_{1}^{{}},\mathbf{r}_{2}^{{}},\mathbf{r}_{1}^{\prime}%
,\mathbf{r}_{2}^{\prime})=-~G~\delta({{\mathbf{R}}}-{{\mathbf{R}}}^{\prime
})~P(r)P(r^{\prime})~{\scriptstyle\frac{1}{2}}(1-P^{\sigma}) \label{vr}%
\end{equation}
where ${{\mathbf{R}}}=\frac{1}{2}({{\mathbf{r}}}_{1}+{{\mathbf{r}}}_{2})$ and
${{\mathbf{r}}}={{\mathbf{r}}}_{1}-{{\mathbf{r}}}_{2}$ are center of mass and
relative coordinates respectively, and $P(r)$ is obtained from the Fourier
transform of $p(k)$. Using the Gaussian ansatz (\ref{gauss-k}) we find%
\begin{equation}
P(r)=\frac{1}{(4\pi a^{2})^{3/2}}e^{-\frac{r^{2}}{4a^{2}}}. \label{gauss-r}%
\end{equation}
Because of the $\delta$-term in Eq.~(\ref{vr}) that insures translational
invariance this force is not completely separable in coordinate space.
However, the matrix elements of this force can be represented by a sum of a
few separable terms in a basis of spherical harmonic oscillator functions.

In order to show this we start from the basis
\begin{equation}
|nljm_{j}\rangle=\varphi_{nljm_{j}}({{\mathbf{r}}})=R_{nl}(r,b)[Y_{l}(\hat
{r})\otimes\chi_{\frac{1}{2}}]_{jm_{j}} \label{XXX}%
\end{equation}
where $R_{nl}(r,b)=b^{-\frac{3}{2}}R_{nl}(r/b)$ and $[Y_{l}(\hat{r}%
)\otimes\chi_{\frac{1}{2}}]_{jm_{j}}$ represents the wave function in spin and
angels coupled to total angular momentum $jm_{j}$. The radial wave function
has the form%
\begin{equation}
R_{nl}(x)=\sqrt{\frac{2n!}{(n+l+\frac{1}{2})!}}x^{l}L_{n}^{l+\frac{1}{2}%
}(x^{2})e^{-\frac{x^{2}}{2}}%
\end{equation}
with the radial quantum number $n=0,1,\dots$ and the orbital angular momentum
$l$. The quantity $b=\sqrt{\hbar/(m\omega_{0})}$ is the harmonic oscillator
length. In the pairing channel we need only the two-particle wave functions
coupled to angular momentum $J=0$ and the projector $\frac{1}{2}(1-P^{\sigma
})$ restricts us to the quantum numbers $S=L=0.$ Recoupling from the $LS$- to
the $jj$-scheme therefore leads to the two-particle wave function%
\begin{align}
|12\rangle_{0}  &  \equiv|\varphi_{n_{1}l_{1}j_{1}}({{\mathbf{r}}}%
_{1}),\varphi_{n_{2}l_{2}j_{2}}({{\mathbf{r}}}_{2})\rangle_{J=0}\label{xxx2}\\
&  =\frac{\hat{\jmath}}{\hat{s}\hat{l}}R_{n_{1}l_{1}}(r_{1},b)R_{n_{2}l_{2}%
}(r_{2},b)|\lambda=0\rangle|S=0\rangle\nonumber
\end{align}
with $\hat{\jmath}=\sqrt{2j+1}$ and $s=\frac{1}{2}$. The functions
$|\lambda=0\rangle=\left[  Y_{l_{1}}(\hat{r}_{1})\otimes Y_{l_{2}}(\hat{r}%
_{2})\right]  _{0}$ and $|S=0\rangle=[\chi_{\frac{1}{2}}\otimes\chi_{\frac
{1}{2}}]_{0}$ are the angular and spin wave functions coupled to angular
momentum $\lambda=0$ and spin $S=0$. All these wave functions are expressed in
laboratory coordinates, while the separable pairing interaction in
Eq.~(\ref{vr}) is expressed in the center of mass frame by the center of mass
coordinate ${{\mathbf{R}}}$ and the relative coordinates ${{\mathbf{r}}}$ of a
pair. Therefore we transform to the center of mass frame using Talmi-Moshinsky
brackets~\cite{Tal.52,Mos.59,BJM.60}. We use the definition of
Baranger~\cite{BD.66}

\begin{figure}[ptb]
\includegraphics[scale=0.26]{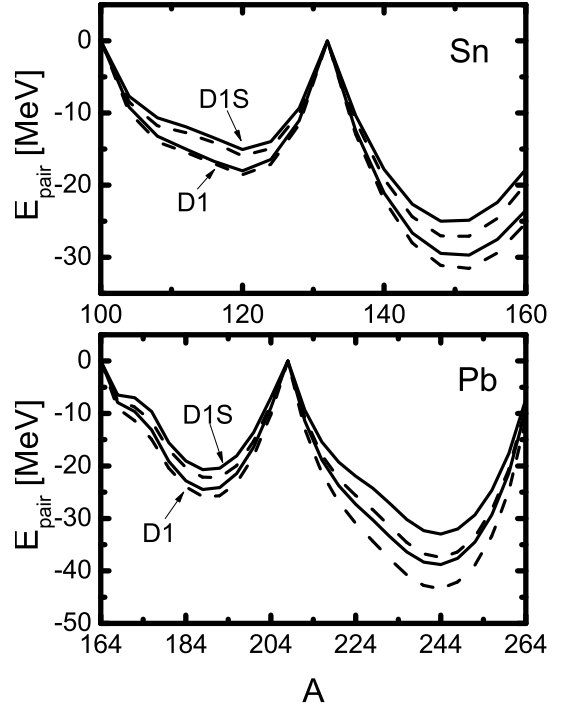}
\caption{Results of RHB calculations in finite nuclei for Sn isotopes
(upper panel) and for Pb isotopes (lower panel): pairing energies for
the Gogny forces D1 and D1S (solid curves) and their corresponding
separable forms
(dashed curves).}%
\label{fig3}%
\end{figure}

\begin{equation}
|n_{1}l_{1},n_{2}l_{2};\lambda\mu\rangle=\sum_{NLnl}M_{n_{1}l_{1}n_{2}l_{2}%
}^{NLnl}|NL,nl;\lambda\mu\rangle\label{xxx3}%
\end{equation}
where%
\begin{equation}
M_{n_{1}l_{1}n_{2}l_{2}}^{NLnl}=\langle NL,nl,\lambda|n_{1}l_{1},n_{2}%
l_{2},\lambda\rangle
\end{equation}
are the Talmi-Moshinsky brackets with the selection rule%
\begin{equation}
2N+L+2n+l=2n_{1}+l_{1}+2n_{2}+l_{2}. \label{n-selection}%
\end{equation}
Here we need these bracket only for the case $\lambda=0$. We therefore can
express the wave function (\ref{xxx2}) in terms of center of mass and relative
coordinates
\begin{align}
|12\rangle_{0}  &  =\frac{\hat{\jmath}}{\hat{s}\hat{l}}\sum_{NL}\sum
_{nl}M_{n_{1}l_{1}n_{2}l_{2}}^{NLnl}\label{xxx4}\\
&  ~~~~~\times R_{NL}(R,b_{R})R_{nl}(r,b_{r})|\lambda=0\rangle|S=0\rangle
\nonumber
\end{align}
with $|\lambda=0\rangle=\left[  Y_{L}(\hat{R})\otimes Y_{l}(\hat{r})\right]
_{\lambda=0}$. The oscillator parameters for the center of mass and the
relative coordinates are $b_{R}=b/\sqrt{2}$ and $b_{r}=b\sqrt{2}$. Finally we
find the pairing matrix elements of the interaction (\ref{vr})
\begin{equation}
V_{121^{\prime}2^{\prime}}^{0}=\langle n_{1}l_{1}j_{1},n_{2}l_{2}%
j_{2}|V|n_{1^{\prime}}l_{1^{\prime}}j_{1^{\prime}},n_{2^{\prime}}l_{2^{\prime
}}j_{2^{\prime}}\rangle_{J=0} \label{ME}%
\end{equation}
as a sum over the quantum numbers $N,$\ $L,$ $N^{\prime}$, $L^{\prime}$, $n$,
$l$, $n^{\prime}$, and $l^{\prime}$ in Eq.~(\ref{xxx3}). The integration over
the center of mass coordinates $\mathbf{R}$ and $\mathbf{R}^{\prime}$ leads to
$N=N^{\prime}$, $L=L^{\prime}$. Further restrictions occur through the fact
that the sum contains integrals over the relative coordinates of the form%
\begin{equation}
\int R_{nl}(r,b_{r})Y_{lm}(\hat{r})P(r)d^{3}r.
\end{equation}
They vanish for $l\neq0$ and this leads to $L=l=0$. The quantum
numbers $n$ and $n^{\prime}$ are determined by the selection rule
(\ref{n-selection}) and we are left with a single sum of separable
terms
\begin{equation}
V_{121^{\prime}2^{\prime}}^{0}=-~G\sum_{N}V_{1^{{}}2}^{N}V_{1^{\prime
}2^{\prime}}^{N} \label{vs}%
\end{equation}
with the single particle matrix elements $V_{12}^{N}$. For $l_{1}=l_{2}=l$,
$j_{1}=j_{2}=j$ we find
\begin{equation}
V_{12}^{N}=M_{n_{1}ln_{2}l}^{N0n0}\frac{\hat{\jmath}}{\hat{s}\hat{l}}%
{\displaystyle\int\limits_{0}^{\infty}}
R_{n0}(r,b_{r})P(r)r^{2}dr.
\end{equation}
For a Gaussian ansatz of $P(r)$ in Eq. (\ref{gauss-r}) this integral can be
evaluated analytically%
\begin{equation}
V_{12}^{N}=\frac{1}{b^{3/2}}\frac{2^{1/4}}{\pi^{3/4}}\frac{(1-\alpha^{2}%
)^{n}\text{ \ \ \ \ }}{(1+\alpha^{2})^{n+3/2}}\frac{\hat{\jmath}}{\hat{s}%
\hat{l}}M_{n_{1}ln_{2}l}^{N0n0}~\frac{\sqrt{(2n+1)!}}{2^{n+1}n!}%
\end{equation}
where the parameter $\alpha=a/b$ characterizes the width of the function
$p(r)$ in units of the oscillator length $b$ and $n$ is given by the selection
rule (\ref{n-selection}) $n=n_{1}+n_{2}+l-N$.

\begin{figure}[ptb]
\includegraphics[scale=0.25]{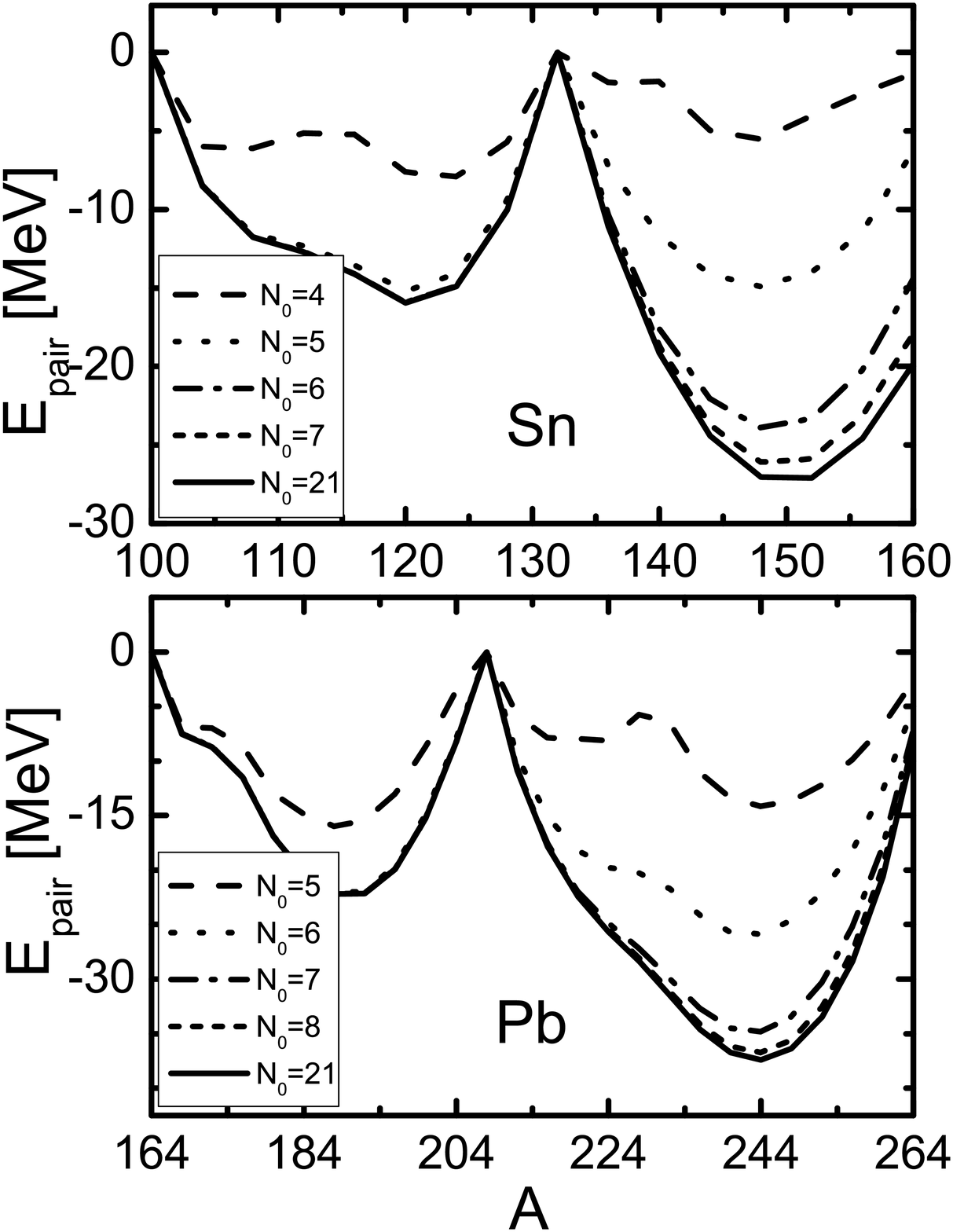}\caption{Pairing energies of Sn- and
Pb-isotopes obtained with different numbers of separable terms
$N_{0}$ in
Eq.~(\ref{vs}).}%
\label{fig4}%
\end{figure}

Thus we find, that the pairing matrix elements for the separable
pairing interactions used in the RHB equation can be evaluated by the
sum of separable terms in Eq.~(\ref{vs}). In order to study the
pairing properties in finite
nuclei, we solve the RHB equation%
\begin{equation}
\left(
\begin{array}
[c]{cc}%
h_{D}-\mu & \Delta\\
\Delta & -h_{D}+\mu
\end{array}
\right)  \left(
\begin{array}
[c]{c}%
U\\
V
\end{array}
\right)  _{k}=E_{k}\left(
\begin{array}
[c]{c}%
U\\
V
\end{array}
\right)  _{k}%
\end{equation}
self-consistently for the Dirac Hamiltonian $h_{D}$ and the pairing field
\begin{equation}
\Delta_{12}=G\sum_{N}P_{N}^{{}}V_{12}^{N},\text{ \ } \label{E41}%
\end{equation}
with the parameters%
\begin{equation}
P_{N}=\frac{1}{2}\sum_{12}V_{12}^{N}\kappa_{12}^{{}}=\frac{1}{2}%
\text{Tr}(V^{N}\kappa)
\end{equation}
and the pairing tensor $\kappa=UV^{T}$. The pairing energy in the nuclear
ground state is given by
\begin{equation}
E_{\text{pair}}=-~G~\sum_{N}P_{N}^{\ast}P_{N}^{{}}. \label{xxx6}%
\end{equation}
All the following calculations are carried out for the parameter set NL3
~\cite{NL3} by expanding the Dirac-Bogoliubov spinors in terms of 20 major
oscillator shells \cite{GRT.90}.

In Fig.~\ref{fig3} we show the dependence of the pairing energy on the neutron
number for the chain of the isotopes $^{100}$Sn $\sim$ $^{160}$Sn and $^{164}%
$Pb $\sim$ $^{264}$Pb. As we see from the upper part of Fig.~\ref{fig3} good
agreement is observed for the pairing energies calculated with the Gogny
pairing force and its separable approximation. The largest discrepancy is less
than 10\%. From this comparison we can conclude that the separable pairing
interaction can describe the paring properties of finite nuclei on almost the
same footing as its corresponding pairing interaction. Therefore, we can use
the separable pairing interaction instead of its complicated original form. In
fact, one could get even better agreement by using the ansatz (\ref{gauss3})
with three separable terms which produces in momentum space identical results
as the Gogny force (see Fig.~\ref{fig1}b).

As we see from the Eq.~(\ref{vs}) the separable pairing interaction is not
fully separable in the spherical harmonic oscillator basis. We have a sum over
the quantum number $N$ characterizing the major shells of the harmonic
oscillator in the center of mass coordinate. In practical applications it
turns out that this sum can be restricted to finite values $N\leq N_{0}=8.$ It
is therefore enough to determine the $N_{0}+1$ matrices $V_{12}^{N}$ at the
beginning of the iteration and to re-calculate the quantities $P_{N}$ in
Eq.~(\ref{E41}) in each step of the iteration. As compared to calculations
with the full Gogny force in the pairing channel this means a considerable
reduction in memory and computer time.

In order to study the convergence with the number of separable terms $N_{0}$
we show in Fig.~\ref{fig4} the pairing energies in the isotopic chains with
$Z=50$ (Sn) and $Z=82$ (Pb) for various values of $N_{0}$. We find that for
the nuclei around the line of $\beta$-stability, $N_{0}=5$ is large enough to
get the full pairing energy; but for the nuclei far from the $\beta$-stability
line, we need at least $N_{0}=8$.

\begin{figure}[ptb]
\includegraphics[scale=0.39]{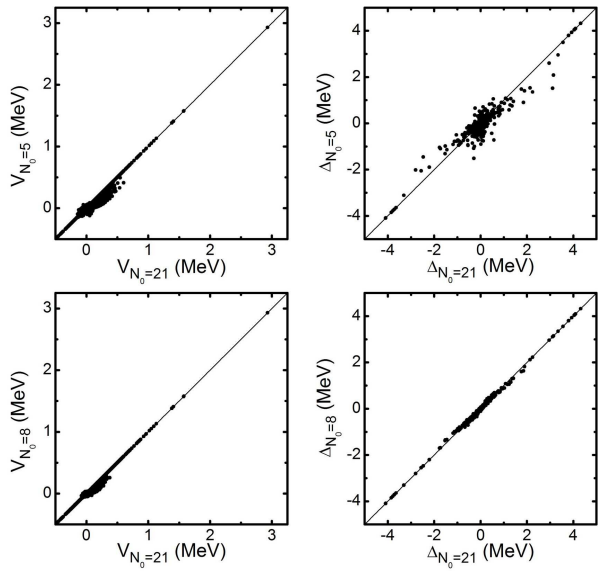}\caption{The pairing matrix elements and
matrix elements of the pairing potential $\Delta$ for the nucleus of $^{244}%
$Pb. The results for various values of $N_{0}$ are compared with
those for $N_{0}=21$ which are identical those obtained with the full
sum in
Eq.~(\ref{vs}).}%
\label{fig5}%
\end{figure}

The left panels of Fig.~\ref{fig5} show the size of the individual matrix
elements (\ref{vs}) calculated with a value $N_{0}=5$ and $N_{0}=8$ as a
function of their exact value (which is identical to the value at $N_{0}=21$).
We see that in particular the large matrix elements are very concentrated
along the 45$^{o}$ line. Only for the very small pairing matrix elements we
find small deviations. Similar results are obtained for the matrix elements of
the pairing field $\Delta_{12}$ of the nucleus $^{244}$Pb. Here we find some
deviations for $N_{0}=5$. However $N_{0}=8$ gives satisfactory agreement.

\begin{figure*}[ptb]
\includegraphics[scale=0.7]{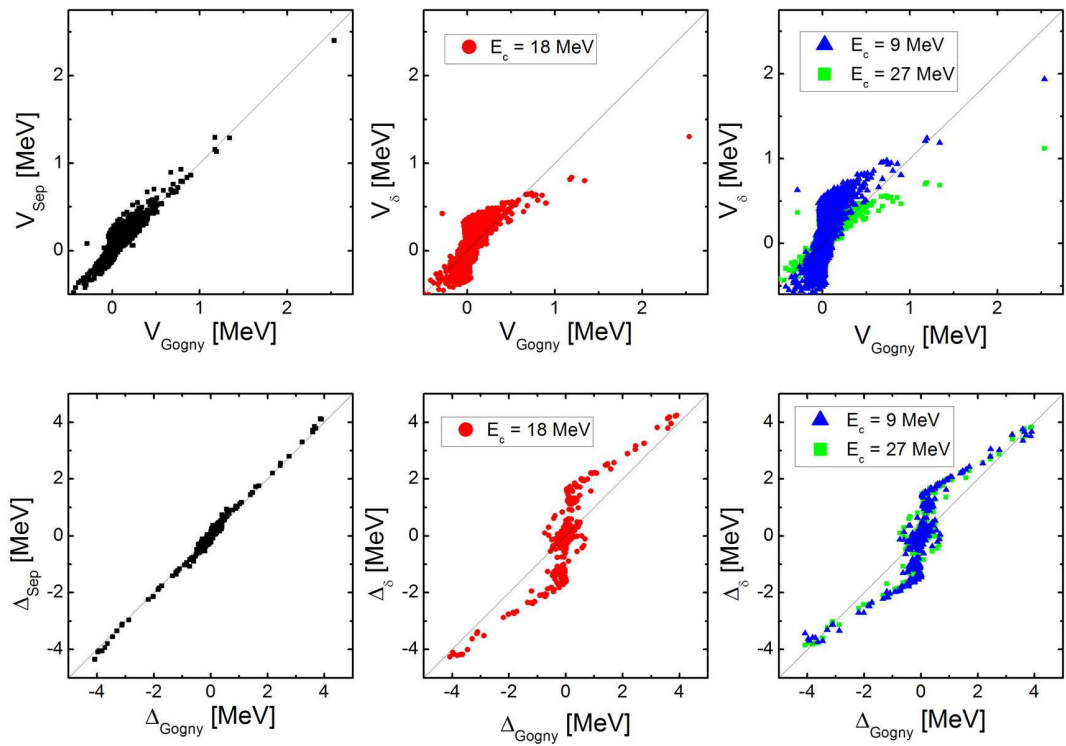}\caption{(color online) Upper panels: The
pairing matrix elements of our separable force and of various
$\delta$-forces with different cut-off parameters $E_{c}$ are
compared with those of the Gogny force D1S. Lower panels: the matrix
elements of the pairing potential $\Delta$ for the nucleus $^{244}$Pb
calculated with the forces shown in the upper panels are compared
with those obtained with the Gogny force D1S. Further
details are given in the text.}%
\label{fig6}%
\end{figure*}

Finally we have carried out a comparison of the new separable pairing force in
Eq.~(\ref{vs}) with the full Brink-Booker part of the Gogny force in the
pairing channel (called full Gogny force in the following), and with various
zero range forces. In the upper panels of Fig.~\ref{fig6} we show the size of
the corresponding pairing matrix elements as a function of the matrix elements
of the full Gogny force with the parameter set D1S~\cite{D1S} and the lower
panels present matrix elements of the pairing field $\Delta_{12}$ in the
oscillator basis for the finite nucleus $^{244}$Pb as a function of those
calculated with the full Gogny force D1S. In the two panels on the left side
we have used the separable pairing force with $N_{0}=21$ which is identical to
the full sum of Eq.~(\ref{vs}). The rest of the panels in Fig.~\ref{fig6} show
calculations with three zero range forces~(\ref{E1}) of various strength
parameters. They have been determined in calculation for the finite nucleus
$^{244}$Pb shown in the lower two panels, where a cut-off energy $E_{c}$ has
been used for the zero range forces. It has been smoothed by a Fermi function
\begin{equation}
f(E) = \frac{1}{1+\exp((E-E_{c})/D_{c})}%
\end{equation}
as in Ref. \cite{BRR.00b}. The smoothing parameter has been kept constant
$D_{c}=0.5$ MeV in all cases and three different values have been chosen for
the cut-off energy, $E_{c}=18$ MeV in the lower middle panel and $E_{c}=9,27$
MeV in the lower right panel. For each cut-off energy the value of the
strength parameter $V_{0}$ has been chosen in such a way that the resulting
average paring gap in the canonical basis
\begin{equation}
\Delta_{av} = \frac{1}{N}\sum_{\mu}|\Delta_{\mu\bar{\mu}}|v_{\mu}^{2}%
\end{equation}
is equal to $\Delta_{av}=1.76$ MeV, the corresponding average gap calculated
with the full Gogny force D1S used in the abscissa of the three lower curves.
We obtain the values $V_{0}=485,326$, and 280 MeV$\cdot$fm$^{3}$ for
$E_{c}=9,18$, and 27 MeV respectively.

We find that although the $\delta$-force can give the same average gap as the
Gogny force D1S if the size of the strength is adjusted properly, the
individual matrix elements of the forces and the matrix elements $\Delta$ of
the pairing field are very different from each other. Apart from many rather
small matrix elements the rest of the matrix elements of the $\delta$-force
cluster around rather constant values. Therefore the $\delta$-force behaves
very much like a constant pairing force with a plateau. As seen in the two
right lower panels of Fig.~\ref{fig6} the value of this plateau depends on the
cut-off energy. On the other side our separable force concentrates along the
45$^{\circ}$ line especially for the large matrix elements. There are only
small differences observed in the region of small pairing matrix elements. For
the pairing potential $\Delta$ we obtain a similar results in the lower panel
of Fig.~\ref{fig6}. Here it is clearly seen that the separable approximation
is very similar to the full Gogny force.

Summarizing, we discuss in this investigation a very simple effective
pairing interaction in the $^{1}S_{0}$-channel, which is of finite
range, translational invariant and separable. This simple force can
be easily applied in realistic applications of modern relativistic
and non-relativistic density functional theory, in particular also in
complicated calculations, such as for nuclei with triaxial shapes,
for nuclei in the rotating frame, for the fission process, for QRPA
calculations and for all kinds of investigations beyond mean field
theory using techniques of projection, generator coordinates, or
particle vibrational coupling. Investigations in this direction are
in progress.

\bigskip
\bigskip

\begin{acknowledgments}
This research has been supported by the National Natural Science Foundation of
China under Grant Nos 10875150, 10775183 and 10535010, the Major State Basis
Research Development of China under contract number 2007CB815000, the
Asia-Europe Link Project [CN/ASIA-LINK/008 (094-791)], the Bundesministerium
f\"{u}r Bildung und Forschung, Germany under project 06 MT 246 and the DFG
cluster of excellence \textquotedblleft Origin and Structure of the
Universe\textquotedblright\ (www.universe-cluster.de).
\end{acknowledgments}


\end{document}